\documentclass[submission,copyright,creativecommons]{eptcs}
\providecommand{\event}{PLACES 2013}
\usepackage[T1]{fontenc}
\usepackage{graphicx}
\usepackage{listings}
\usepackage{color}

\definecolor{darkviolet}{rgb}{0.5,0,0.4}
\definecolor{darkgreen}{rgb}{0,0.4,0.2} 
\definecolor{darkblue}{rgb}{0.1,0.1,0.9}
\definecolor{darkgrey}{rgb}{0.5,0.5,0.5}
\definecolor{lightblue}{rgb}{0.4,0.4,1}

\lstdefinestyle{eclipse}{
  breaklines=true,
   emphstyle=\color{red}\bfseries, 
   keywordstyle=\color{darkviolet}\bfseries,
   commentstyle=\color{darkgreen},
   stringstyle=\color{darkblue},
   numberstyle=\color{darkgrey},
   emphstyle=\color{red},
   morecomment=[s][\color{lightblue}]{/**}{*/},
  showstringspaces=false
}

\usepackage[scaled]{beramono}
\newcommand\Small{\fontsize{8.2}{8.4}\selectfont}

\newcommand{\ccode}[1]{\lstinline[language=C,style=eclipse]@#1@}

\lstdefinelanguage{protocol}{
  basicstyle=\Small\ttfamily,
  style=eclipse,
  keywords={loop,choice,end,message,broadcast,foreach,scatter,gather,allgather,reduce,allreduce,float,int},
  morekeywords={send,receive},  
  morekeywords={nat},  
  morekeywords={max,min,sum,prod,nand,land,band,lor,bor,lxor,bxoer,minloc,maxloc},  
  morecomment=[l]{//}, 
  morecomment=[s]{/*}{*/}, 
  morestring=[b]",
  tabsize=2,
  literate=
    {Pi}{$\Pi$}1
}

\newcommand{\commcode}[1]{\lstinline[language=protocol,style=eclipse]@#1@}

\newcommand{\lasige}{LaSIGE, Faculty of Sciences\\University of Lisbon, Portugal}
\title{Towards deductive verification of MPI programs against session types}
\author{
  Eduardo R. B. Marques
  \institute{\lasige}
  \and
  Francisco Martins
  \institute{\lasige}
  \and
  Vasco T.\ Vasconcelos 
  \institute{\lasige}
  \and
  Nicholas Ng
  \institute{ Imperial College London, UK}
  \and
  Nuno Martins
  \institute{\lasige}
}
\def\titlerunning{Towards deductive verification of MPI programs against session types}
\def\authorrunning{Marques, Martins, Vasconcelos, Ng, and Martins}

\begin{document}

\maketitle
\begin{abstract}
  The Message Passing Interface (MPI) is the de facto standard
  message-passing infrastructure for developing parallel
  applications. Two decades after the first version of the library
  specification, MPI-based applications are nowadays routinely
  deployed on super and cluster computers.
  These applications, written in C or Fortran, exhibit intricate message
  passing behaviours, making it hard to statically verify important
  properties such as the absence of deadlocks.
  Our work builds on session types, a theory for describing
  protocols that provides for correct-by-construction
  guarantees in this regard.  We annotate MPI primitives and C code with session type
  contracts, written in the language of a software verifier for C.
  Annotated code is then checked for correctness with the software
  verifier.
  We present preliminary results and discuss the challenges that lie
  ahead for verifying realistic MPI program compliance against session
  types.
\end{abstract}


\section{Introduction}


MPI is a library specification targeting the development of
communication intensive parallel applications~\cite{mpi3}.  There are
a number of libraries available that allow to use MPI primitives from
within C or Fortran code.  MPI supports a huge collection of
communication primitives, including collective (barrier, broadcast,
reduction, \dots) and point-to-point communications (blocking and non
blocking), supports persistence, datatypes (predefined and user
defined), communication contexts, different process topologies,
one-sided communications, file I/O, among others.
%


Communication mismatch is a major source of errors in MPI
applications, often leading to deadlocks or to erroneous results.
However, verification of MPI applications is non-trivial, accounting
for an active research area. A recent survey \cite{cacm} summarises
the state of the art of verification methods for MPI. Most of these
methods are based on model-checking.
ISP~\cite{isp-ppopp09} is a dynamic verifier which uses a scheduler to explore
all possible thread interleavings of an execution. The tool exploits
independence between thread actions as a heuristic to avoid state explosion. A
fixed test harness is then used to detect common deadlock patterns.  While this
traditional model checking technique aims to capture most common deadlock
patterns in MPI programs, their approach is limited to a finite number of
tests, and remains to be an approximate solution for deadlock detection.
TASS~\cite{tass-ppopp11} is a tool that combines symbolic execution
\cite{siegel-tosem08} and model checking techniques to verify safety properties
of MPI programs. The tool takes a C+MPI application and an input $n \ge 1$
which restricts the input space, then constructs an abstract model with $n$
processes and checks its functional equivalence with a sequential
implementation by executing the model of the application.
Parallel data-flow analysis is a static analysis technique applied in
\cite{dfa-cgo09}. The work focuses on send-receive matching in MPI source
code, which helps identify message leaks and communication mismatch, by
constructing a parallel control-flow graph by simple symbolic analysis on the
control-flow graph of the MPI program. In \cite{dfa-ics13} the authors
discuss extending the technique by combining static and dynamic analysis to
improve precision of the data-flow analysis.


Model-checking based MPI verification methods rely on external testing,
or equivalence checking of the model against a correct sample implementation.
Parallel data-flow analysis is somewhat ad-hoc.
In contrast we seek a simple constructive verification method for an overall
understanding of the communication patterns of programs. Our method
aims at ensuring that programs follow a predefined protocol; in the
process, we ensure that programs are exempt from communication
mismatches, deadlocked situations in particular.


\begin{figure}[t]
  \centering
  \includegraphics[width=0.66\textwidth]{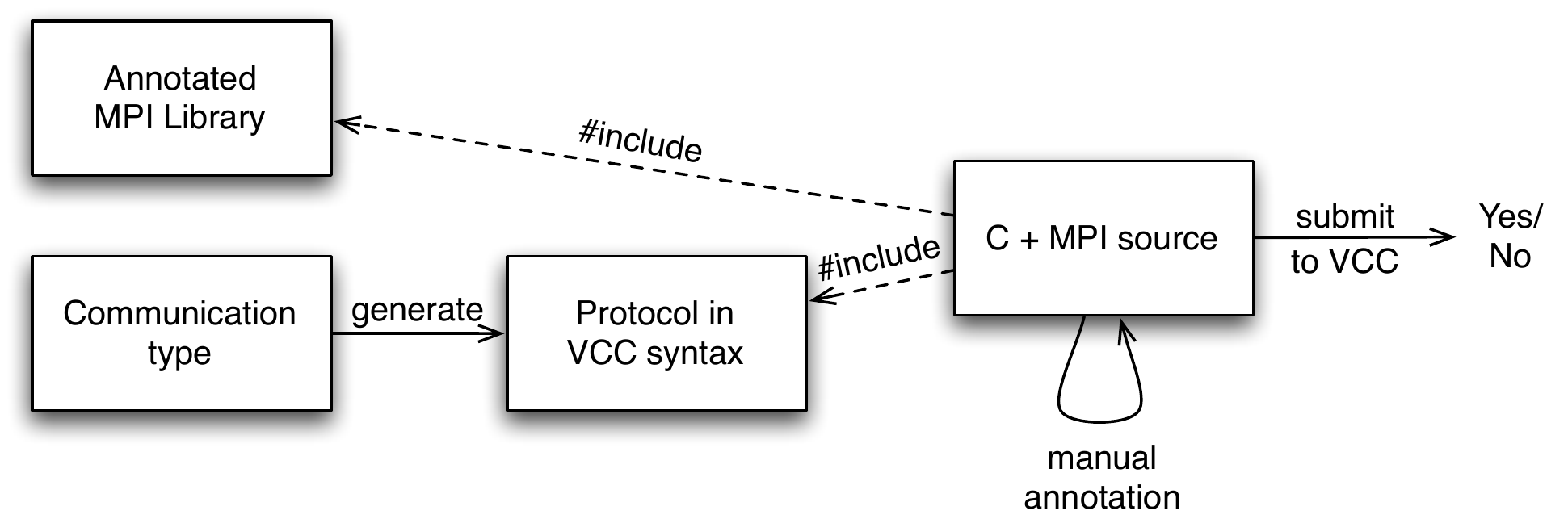}
  \caption{Outline of the approach}
\label{fig:approach}
\end{figure}


Our approach is depicted in Figure~\ref{fig:approach}. We start by writing the protocol in a language
tailored for describing MPI communication patterns. Afterwards, we
translate the protocol into a term written in the language of a
software verifier tool for the C programming language, VCC~\cite{vcc}.
The C+MPI code imports the protocol definition (in VCC form) and a
VCC-annotated MPI header with session type contracts for the various
MPI primitives. Depending on the specifics of the C code, further
manual annotations may be required. In this setting, VCC is invoked to
check whether the C code follows the communication type. 


The project closest to ours is Scribble~\cite{scribble}. Based on the
theory of Multiparty Session Types~\cite{Multiparty-POPL}, Scribble
describes, from an high level perspective, patterns of message-passing
interactions.  Protocol design with Scribble starts by identifying the
communication participants.  The body of the protocol describes the
interactions from a global viewpoint, with explicit senders and
receivers, thus ensuring that all senders have a matching receiver and
vice versa. Global protocols are then projected into each of their
participants' counterparts, yielding one local protocol for each
participant present in the global protocol. The projection algorithm
converts the user-defined global protocol into the local interaction
behaviour of each participant automatically, while preserving the
overall interaction patterns and the order of the interactions.
Developers can then implement programs for the various individual
participants, based on the local protocols and using standard
message-passing libraries.
One such approach, Multiparty Session-C, builds a library of session
primitives to be used within the C language~\cite{session_c}.


In this work we slightly depart from Multiparty Session Types and
Scribble by introducing collective decision primitives, allowing for
behaviours where all participants decide to enter or to leave a loop, 
or choose one of the two branches of a choice point, two patterns 
impossible to describe with Scribble. We found these
primitives to be in line with the common practice of MPI programming.
We have also included in the language of communication types MPI
specific collective operations, as well as a dependent functional type
constructor. Finally, and in contrast to Session-C where programmers
use a particular library of communication operations, we directly
check standard C+MPI code. Preliminary ideas were put forward
in~\cite{eurompi}.


The outline of the this paper is as follows. The next section
introduces our running example, written in C+MPI, as well as the
language of communication types. Section~\ref{sec:verifying}
explains the process of verifying C+MPI code against communication
types, and Section~\ref{sec:results} presents results of running our
system on four textbook examples. Section~\ref{sec:conclusion}
concludes the paper and points a few directions for further work.


\section{Communication types for MPI programs}


\begin{figure}[t]
\begin{lstlisting}[numbers=left] 
int main(int argc, char** argv) {
  int np;                          // Number of processes
  int me;                          // My process rank
  MPI_Init(&argc, &argv);
  MPI_Comm_size(MPI_COMM_WORLD, &np);
  MPI_Comm_rank(MPI_COMM_WORLD, &me);
  ...
  int psize = atoi(argv[1]);       // Global problem size
  if (rank == 0)
    read_vector(work, psize);
  ...
  // Scatter input data
  MPI_Scatter(work, lsize, MPI_FLOAT, &local[1], lsize, MPI_FLOAT, 0, MPI_COMM_WORLD);
  ...
  int left  = (np + me - 1) % np;  // Left neighbour
  int right = (me + 1) % np;       // Right neighbour
  // Loop until finite differences converge to a minimum error or max iterations attained
  while (!converged(globalerr) && iter < MAX_ITER) {
    ...
    if (me % 2 == 0) {
      MPI_Send(&local[1],       1, MPI_FLOAT, left,  0, MPI_COMM_WORLD);
      MPI_Recv(&local[lsize+1], 1, MPI_FLOAT, right, 0, MPI_COMM_WORLD, &status);
      MPI_Recv(&local[0],       1, MPI_FLOAT, left,  0, MPI_COMM_WORLD, &status);
      MPI_Send(&local[lsize],   1, MPI_FLOAT, right, 0, MPI_COMM_WORLD);
    } else {
      MPI_Recv(&local[lsize+1], 1, MPI_FLOAT, right, 0, MPI_COMM_WORLD, &status);
      MPI_Send(&local[1],       1, MPI_FLOAT, left,  0, MPI_COMM_WORLD);
      MPI_Send(&local[lsize],   1, MPI_FLOAT, right, 0, MPI_COMM_WORLD);
      MPI_Recv(&local[0],       1, MPI_FLOAT, left,  0, MPI_COMM_WORLD, &status);
    }
    ...
    MPI_Allreduce(&localerr, &globalerr, 1, MPI_FLOAT, MPI_MAX, MPI_COMM_WORLD); 
    ...
  } 
  ...
  if (converged(globalerr)) {
    // Gather data at rank 0 for solution
    MPI_Gather(&local[1], lsize, MPI_FLOAT, work, lsize, MPI_FLOAT, 0, MPI_COMM_WORLD);
    ...
  } else
    printf ("Failed to converge after %d iterations!", MAX_ITER);
  MPI_Finalize();
  return 0; 
}
\end{lstlisting} 
\caption{Excerpt of an MPI program for the finite differences
  algorithm (adapted from~\cite{foster})}
\label{fig:running-example}
\end{figure}


Our running example is that of the one-dimensional finite differences
problem, in which we start with a vector~$X^0$ and must compute $X^T$
iteratively, governed by a given recurrence formula~\cite{foster}. The
original C+MPI code is given in Figure~\ref{fig:running-example}.
Even though the same program is executed by all processes (in line
with the Single-Program-Multiple-Data paradigm), each process is
endowed with a unique natural number, its \emph{rank}, that can be
used to partially specialise its behaviour.

MPI programs start with a call to \lstinline|MPI_Init| and conclude
with a call to \lstinline|MPI_Finalize|, lines 4 and 42 in
Figure~\ref{fig:running-example}.  After initialising the framework,
each process asks for the total number of processes
(\lstinline|MPI_Comm_size|, line 5) and the process' own rank
(\lstinline|MPI_Comm_rank|, line 6), storing these values in variables
\lstinline|np| and \lstinline|me|, respectively. The problem size is
read, by all processes, from the arguments of the program and into
variable \lstinline|psize| (line 8).
The process with rank \lstinline|0| alone reads vector $X^0$ into
memory (lines 9--10), and then distributes it among all participants
(including itself), each participant receiving a slice of length
\lstinline|psize/np| elements (\lstinline|MPI_Scatter|, line 13).

Each process then loops until the finite differences converges to a
given threshold or a given number of iterations is attained, lines
18--34. The body of the loop specifies point-to-point message
exchanges (\lstinline|MPI_Send| and \lstinline|MPI_Recv|) between each
process and its left and right neighbours, following a ring
topology. The purpose of these exchanges is to distribute the border
values necessary for the calculations due to each participant.
We assume the standard \emph{non-buffered, synchronous semantics} of
MPI operations, in that, e.g., an \lstinline{MPI_Send} operation blocks until
the target process issues the corresponding \lstinline{MPI_Recv}
operation.
This justifies the different orderings of the message exchanges for the
even and the odd ranked participants (lines 20--24 and 26--29): processes would
deadlock otherwise.
After the two (per participant) message exchanges, and before the end
of the loop step, the global error is calculated with a reduction
operation and propagated to all participants
(\lstinline|MPI_Allreduce|, line 32).
If the procedure converges, the solution is gathered at
rank~\lstinline|0| (\lstinline|MPI_Gather|, line 38).


\begin{figure}[t]
  \begin{minipage}[t]{0.33\textwidth}
    \begin{lstlisting}[language=protocol,numbers=left]
Pi size: {n:nat|n%3==0}. 
scatter(0,MPI_FLOAT,size/3). 
loop(
  message(2,1,MPI_FLOAT,1).
  message(0,2,MPI_FLOAT,1).
  message(1,0,MPI_FLOAT,1).
  message(1,2,MPI_FLOAT,1).
  message(2,0,MPI_FLOAT,1).
  message(0,1,MPI_FLOAT,1).
  allreduce(MPI_FLOAT,1,MPI_MAX). 
  end).
choice(
  gather(0,MPI_FLOAT,size/3).end,
  end). 
end
    \end{lstlisting}
  \end{minipage}
  \begin{minipage}[t]{0.33\textwidth}
    \begin{lstlisting}[language=protocol]
Pi size: {n:nat|n%3==0}.
scatter(0,MPI_FLOAT,size/3).
loop(
  send(2,MPI_FLOAT,1).
  receive(1,MPI_FLOAT,1).
  receive(2,MPI_FLOAT,1).
  send(1,MPI_FLOAT,1).
  allreduce(MPI_FLOAT,1,MPI_MAX).
  end).
choice(
  gather(0,MPI_FLOAT,size/3).end,
  end).
end
    \end{lstlisting}
  \end{minipage}
  \begin{minipage}[t]{0.33\textwidth}
    \begin{lstlisting}[language=protocol]
Pi size: {n:nat|n%3==0}.
scatter(0,MPI_FLOAT,size/3).
loop(
  receive(2,MPI_FLOAT,1).
  send(0,MPI_FLOAT,1).
  receive(2,MPI_FLOAT,1).
  send(0,MPI_FLOAT,1).
  allreduce(MPI_FLOAT,1,MPI_MAX).
  end).
choice(
  gather(0,MPI_FLOAT,size/3).end,
  end).
end
    \end{lstlisting}
  \end{minipage}
  \caption{Global communication type and two local types (ranks 0 and
    1)}
  \label{fig:type-example}
\end{figure}


The overall protocol for the various processes is described as a
term in the language of \emph{communication types}. Such a term
captures, not only the various message exchanges between processes
(point-to-point, broadcast), but also the
communication-related loops and choices programs make. 
We informally describe the language next. 
A possible communication type for
our running example is presented in Figure~\ref{fig:type-example},
left column.

The atoms in our types describe the individual MPI communications and
the special dependent function type constructor, \commcode{Pi}.
Line 2, \commcode{scatter(0,MPI_FLOAT,size/3)},
describes a data distribution operation, initiated at rank
\commcode{0} and delivering a float array of length \commcode{size/3}
to each process. The operation in line 13,
\commcode{gather(0,MPI_FLOAT,size/3)}, behaves similarly except that it
gathers at rank \commcode{0} the various slices of an array.
Lines 4--9 introduce point-to-point communications. For example,
\commcode{message(2,1,MPI_FLOAT,1)} describes a message exchange, from
process rank \commcode{2} to process rank \commcode{1}, containing a
float array of length 1.

Individual MPI communications are composed via \emph{prefixing} and
\emph{collective decisions}. Prefixing is defined by the \commcode{.} (dot) operator,
and the terminated protocol is denoted by
\commcode{end}. A protocol that scatters and then terminates can be
written as \commcode{scatter(0,MPI_FLOAT,size/3).end}.
Collective decisions include \emph{loops} and \emph{choices}. A type
\commcode{loop(allreduce(MPI_FLOAT,1,MPI_MAX).end).T} denotes a point
in the protocol where \emph{all} processes either decide to enter or
to leave the loop. In case a process enters the loop, it performs an
\commcode{allreduce} operation; in case it decides not to enter the
loop, it continues as \commcode{T}. The case of a choice is similar: a
type \commcode{choice(gather(0,MPI_FLOAT,size/3).end,end).T} describes
a point in the protocol where \emph{all} processes either decide to
gather a float array, or not to engage in any MPI operation. In either
case, each process then continues as prescribed by \commcode{T}.

Ranks and array lengths are described by integer expressions. The
communication type under discussion mentions variable
\commcode{size}. Such a variable is introduced by a dependent function
type constructor~\commcode{Pi}. A type
\lstinline[language=protocol]@Pi size:{n:nat|n
denotes a protocol parametric on the size of the problem that
scatters a float array in chunks of length \lstinline[language=protocol]@size/3@.
Expressions in communication types are formed from literals, variables
and arithmetic expressions. Such expressions are of kind integer
(\commcode{int}), floating point (\commcode{float}), or array
(\commcode{float[n]}).
Furthermore, any such kind can be \emph{refined}. Kind
\lstinline[language=protocol]@{n:nat|n
in line 1 denotes a non-negative integer, multiple of 3. Kind \commcode{nat} is
itself an abbreviation for
\lstinline[language=protocol]@{n:int|n>=0}@.
%
%
The program in Figure~\ref{fig:running-example} works for any number
of participants. In contrast the communication type in
Figure~\ref{fig:type-example} describes a protocol for exactly three
participants, ranked \commcode{0} to \commcode{2}. The current version
of our communication type language does not allow for iteration, hence 
we hard-coded the
 number of processes (\commcode{3} in the example) 
 in the type.


We can easily see that there is roughly a one-to-one correspondence
between the MPI primitives in our running program and those in the
communication type, including loops and conditionals, except for the
point-to-point communications, where we see \lstinline@MPI_Send@ and
\lstinline@MPI_Recv@ in Figure~\ref{fig:running-example} and
\commcode{message} alone in Figure~\ref{fig:type-example}. There is a
difference in perspective: the C code describes a per-process (or
\emph{local}) view, whereas the communication code presents a
\emph{global} view of the protocol.
In order to check C code against communication types, we have to
reduce the global view of communication types into a local view.
Following~\cite{Multiparty-POPL}, we \emph{project} the communication
type into each of the ranks present in the type, thus obtaining a
series of \emph{local} protocols. Local protocols are very much like
global protocols. The only difference is that a
\commcode{message(0,1,MPI_FLOAT,1)} point-to-point communication is
replaced by \commcode{send(1,MPI_FLOAT,1)} when projecting on rank
\commcode{0}, by \commcode{receive(0,MPI_FLOAT,1)} when projecting on
rank \commcode{1}, and omitted altogether for all other ranks.

The projections of the global protocol on ranks \commcode{0} and
\commcode{1} are shown in the central and right columns of
Figure~\ref{fig:type-example}. We can easily see that the overall
structure of the protocol is preserved, except for the point-to-point
communications, where the six occurrences of \commcode{message} in the
communication type are replaced by four occurrences in each local type
(notice that each rank \commcode{0}--\commcode{2} occurs four times in
lines 4--9 of the communication type). It worth noting that the
projection operation alone yields different local types for the odd
and for the even ranks, as witnessed by the two local types in
Figure~\ref{fig:type-example}, and that these are aligned with the C
code (Figure~\ref{fig:running-example}, lines 21--24 and 26--29).


\section{Verifying C+MPI code against communication types}
\label{sec:verifying}

\begin{figure}[t]
  \begin{minipage}[t]{0.5\textwidth}
    \begin{lstlisting}
typedef int MPI_Datatype;
typedef int Rank;
typedef int Length;
_(datatype \Comm {
  case send(Rank,MPI_Datatype,Length); 
  case recv(Rank,MPI_Datatype,Length); 
  case scatter(Rank,MPI_Datatype,Length);
  case gather(Rank,MPI_Datatype,Length);
  case bcast(Rank,MPI_Datatype,Length);
  ...
})
_(datatype \Type {
  case end();
  case comm(\Comm, \Type);
  case loop(\Type, \Type);
  case choice(\Type, \Type, \Type);
})
    \end{lstlisting}
  \end{minipage}
  \begin{minipage}[t]{0.5\textwidth}
    \begin{lstlisting}
_(ghost _(pure) \Type type_func(int rank, int size)
  _(requires 0 <= rank && rank < 3)
  _(requires 0 <= size && size % 3 == 0)
  _(ensures \result ==
    comm(scatter(0,MPI_FLOAT,size/3),
    loop (
      rank == 0 ? 
        comm(send(2,MPI_FLOAT,1), ...) :
      rank == 1 ?
        comm(recv(2,MPI_FLOAT,1), ...) :
      // rank == 2
        comm(send(1,MPI_FLOAT,1), ...),
    choice(
      comm(gather(0, MPI_FLOAT, size/3),end()),
      end(),
    end())))))
\end{lstlisting}
  \end{minipage}
  \caption{The VCC datatype for communication types and the type function
    for the running example}
  \label{fig:vcc-example}
\end{figure}



Communication types, as described in the previous section, are
translated into the language of VCC so that they may be used in the
verification process.
The language of communication types is described in VCC by a
datatype. The left column in Figure~\ref{fig:vcc-example} describes
the datatype \lstinline|\Type| for communication types, that for
convenience relies on datatype \lstinline|\Comm| for MPI operations.


Rather than translating each local type individually as suggested by
the central and right columns in Figure~\ref{fig:type-example}, the
global type is translated as a VCC function that takes a process rank
as parameter. It is this function, \lstinline@type_func@, that
internally projects \emph{all} ranks (as described in the previous
section), by making use of the conditional expression (\lstinline@?:@)
of the C programming language.
%
%
Communication types can be parametric on program values; one such
example is exhibited in Figure~\ref{fig:type-example}, line 1, where
the problem size is introduced in the type. Currently we allow
dependent function types to occur only at the top level of types. The
parameters for the various dependent functions are all gathered at
\lstinline@type_func@,  in addition to the rank parameter.


The right column in Figure~\ref{fig:vcc-example} contains the VCC
function corresponding to the communication type in
Figure~\ref{fig:type-example}, where we can find two parameters
corresponding to the rank and to the problem size.
Some explanation on the syntactic details of VCC are in order. VCC
annotation blocks are introduced as \lstinline@_(annotation block)@;
keyword \lstinline@ghost@ introduces an annotation block necessary for
the verification process, but inconsequential for the C program;
keyword \lstinline@pure@ describes a function without side effects;
keywords \lstinline@requires@ and \lstinline@ensures@ introduce the
pre and post conditions of a function, respectively. Finally keyword
\lstinline@\result@ denotes the value of the function.


\begin{figure}[t]
\begin{lstlisting}[numbers=left] 
int main(int argc, char** argv _ampi_arg_decl) {
  ...
  MPI_Init(&argc, &argv);
  MPI_Comm_size(MPI_COMM_WORLD, &np);
  MPI_Comm_rank(MPI_COMM_WORLD, &me);
  _(assume np == 3)
  ...
  int psize = atoi(argv[1]);     // Global problem size
  _(ghost type = type_func (me, psize))
  ...
  MPI_Scatter(work, lsize, MPI_FLOAT, &local[1], lsize, MPI_FLOAT, 0, MPI_COMM_WORLD);
  ...
  int left = (np + me - 1) % np; // Left neighbour
  int right = (me + 1) % np;     // Right neighbour
  _(ghost \Type loop_body = loopBody(type);)
  _(ghost \Type loop_continuation = next(type);)
  while (!converged(globalerr) && iter < MAX_ITER) 
    _(writes &globalerr)
    _(writes \array_range(local, (unsigned) lsize + 2))
  {
    _(ghost type = loop_body;)
    ...
    _(assert type == end())
  } 
  _(ghost type = loop_continuation;)
  ...
  _(ghost \Type choice_true = choiceTrue(type);)
  _(ghost \Type choice_false = choiceFalse(type);)
  _(ghost \Type choice_continuation = next(type);)
  if (converged(globalerr)) {
    _(ghost type = choice_true;)  
    MPI_Gather(&local[1], lsize, MPI_FLOAT, work, lsize, MPI_FLOAT, 0, MPI_COMM_WORLD);
    ...
    _(assert type == end())
  } else {
    _(ghost type = choice_false;)
    printf ("failed to converge after %d iterations!", MAX_ITER);
    _(assert type == end())
  }
  _(ghost type = choice_continuation;)
  MPI_Finalize();
  return 0; 
}
\end{lstlisting} 
\caption{Annotated version of Figure~\ref{fig:running-example}}
\label{fig:running-example-annotated}
\end{figure}


As described in Figure~\ref{fig:approach}, function
\lstinline|type_func| is placed in a C header file and included in our
running example. Unfortunately, including header files alone is not
enough to check C+MPI code against communication types. VCC
annotations (some of which can be easily automated) must be added to
the code. Figure~\ref{fig:running-example-annotated} shows the code
for our running example with the necessary annotations.
%
%
The first annotation is in line 1. The \lstinline{_ampi_arg_decl} is a
convenience C preprocessor macro that introduces a number of ghost
parameters used by the verification logic, employed by
\lstinline{main} or any other C function of interest.  The declaration
of \lstinline{main} in line~1 expands to
\begin{lstlisting}
int main(int argc, char** argv _(ghost _ampi_glue_t* _gd) _(ghost \Type _type) _(out \Type _type_out) 
\end{lstlisting}
The ghost parameters specify the input and output session type for a
function (\lstinline{_type} and \lstinline{_type_out}), discussed
later in the text, and the \lstinline{gd} argument characterises
verification data for the overall restrictions on the number of
processes and the process rank. The \lstinline{_ampi_glue_t}
declaration contains data fields and VCC data structure invariants for
the restrictions, as follows:
\begin{lstlisting}
typedef struct {
  int procs;  _(invariant 1 < procs && procs < 32768)
  int rank;   _(invariant 0 <= rank && rank < procs)
} ampi_glue_t;
\end{lstlisting}

The parameterisation of \lstinline{gd->procs} and \lstinline{gd->rank}
are reflected in the contracts of primitives \lstinline{MPI_Comm_size}
and \lstinline{MPI_Comm_rank}:
\begin{lstlisting}
int MPI_Comm_size(MPI_Comm* comm, int* size _(ghost ampi_glue_t* gd))
  _(requires comm == MPI_COMM_WORLD)
  _(ensures *size == gd->procs)
  ...
int MPI_Comm_rank(MPI_Comm* comm, int* rank _(ghost ampi_glue_t* gd))
  _(requires comm == MPI_COMM_WORLD)
  _(ensures *rank == gd->rank)
  ...
\end{lstlisting}
Recall that these primitives are used to obtain the number of
processes and the process rank (lines~4--5 in the example), and note
that we currently restrict the MPI communicator to be only the global
top-level communicator in MPI, that is, \lstinline{MPI_COMM_WORLD}.

After line~1, the annotation of the program is resumed with a (refined) restriction
for the number of processes (\lstinline{_(assume np == 3)} at line 6) and
a ghost call to the \lstinline@type_func@ function at line 9. 
If providing the number of processes to the
function can be easily automated, only the programmer knows which
expression in the program corresponds to the problem size.
%
%
%
Furthermore, rather than annotating calls to MPI primitives at each
call site, a contract is defined for each primitive. These contracts
rely on the \lstinline{first} and the \lstinline{next} partial
functions, both operating on \lstinline{\Type}, and defined by the
following axioms.
\begin{lstlisting}
\forall \Type t; \forall \Comm c; first(comm(c, t)) == c
\forall \Type t; \forall \Comm c; next(comm(c,t)) == t
\forall \Type t1,t2; next(loop(t1,t2)) == t2
\forall \Type t1,t2,t3; next(choice(t1,t2,t3)) == t3
\end{lstlisting}

Using these partial functions, the contract for the
\lstinline{MPI_Send} primitive can be expressed as follows,
\begin{lstlisting}
extern int _MPI_Send (void *buf, int count, MPI_Datatype dtype, int target, int tag,
    MPI_Comm c _(ghost ampi_glue_t* gd) _(ghost \Type _type) _(out \Type _type_out))
  _(requires first(_type) == send(target, dtype, count))
  _(requires dtype == MPI_INT ==> \thread_local_array((int*) buf, count))
  _(requires dtype == MPI_FLOAT ==> \thread_local_array((float*) buf, count))
  _(ensures _type_out == next(_type))
  ...
\end{lstlisting}
where we use two ghost variables, \lstinline{_type} and
\lstinline{_type_out}, to represent the types before and after the call
to \lstinline{MPI_Send}, respectively (VCC does not allow in/out ghost
parameters).
The contract states that the first action of the incoming type must be
of the form \lstinline|send(target,dtype,count)|, where
\lstinline|target| is the rank of the destination process, and
\lstinline{dtype} and \lstinline{count} define the data to be
transmitted in the message.
We must also check the data part of MPI primitives, and this is where
VCC becomes handy. In this case we check that the type of the buffer
array matches the declared MPI type, and that the buffer contains
enough space.
%
For example, \lstinline{\thread_local_array((int*)buf,count)} means
that the memory from \lstinline{&buf[0]} to \lstinline{&buf[count-1]}
is valid and typed as \lstinline{int}.
The post condition expresses the effect of the \lstinline|MPI_Send| operation on
the type: after the send operation, the outgoing type is the incoming
type from which the first communication has been removed.
The remaining MPI primitives have similar contracts.
At the end of the program, that is, at the calls to
\lstinline{MPI_Finalize}, we check that the type has reduced to
\lstinline{end}.
\begin{lstlisting}
extern int _MPI_Finalize(_(ghost \Type _type) _(out \Type _type_out)) 
  _(requires _type == end())
  _(ensures _type_out == end()) 
\end{lstlisting}


For collective operations, loops in particular, we currently follow a
very simple and intentional approach: a loop in the type
(\commcode{loop)} must be matched by a loop in the code
(\lstinline|for| or \lstinline|while|). We require a further (partial)
function to extract the body of a \lstinline{loop} type,
governed by the following axiom.
\begin{lstlisting}
\forall \Type t1,t2; loopBody(loop(t1,t2)) == t1
\end{lstlisting}
Equipped with such a function the main loop of our running example is
annotated as in Figure~\ref{fig:running-example-annotated},
lines~15--25.
%
%
The code is self-explanatory: we extract the loop body and the
continuation types at loop entry (lines 15--16). Then, enter the loop
with type \lstinline|loop_body| and terminate the loop with type
\lstinline|end|. 
In order to analyse the rest of the program we use
the \lstinline|loop_continuation| type (line 25).
The case of a \lstinline{choice} is handled similarly in lines~27--40.
In addition to annotations related to the session type,
others are required by VCC in regard to the use of of memory,
for proper inference of side-effects.
For instance, in lines~18--19, the write clause annotations indicate 
that the variable \lstinline{globalerr} and the array \lstinline{local} (from position \lstinline{0} to \lstinline{lsize+2}) are changed in the  loop body.


VCC analyses C code modularly, each function separately. This means
that each such function needs a contract that, among other things,
describes the communication type at entry and at exit points. Suppose
for example that the reading and distribution of the data among all
processes (lines 9--13 in Figure~\ref{fig:running-example}) is
abstracted in a function \ccode{read_vector}. The (currently) manually
annotated function signature would look as follows:
\begin{lstlisting}
void read_vector(int psize, int me, int np, float local[]
    _(ghost ampi_glue_t* gd) _(ghost \Type _type) _(out \Type _type_out))
  ...
  _(requires psize >= 0)
  _(requires psize % np == 0)
  _(requires first(_type) == scatter(0, MPI_FLOAT, psize/np)) 
  _(ensures _type_out == next(_type))
  ...
;
\end{lstlisting}
The logic is similar to that employed for MPI primitives. 
Ghost variables are used in the function declaration 
to represent the input and output  session type, \lstinline{_type}
and \lstinline{_type_out} respectively, and 
the contract states the actions performed 
within the function. In this case, we have that
the first action of the input type should be 
\lstinline{scatter(0, MPI_FLOAT, psize/np)} and that
the output type is the continuation of the input type.


\section{Results}
\label{sec:results}

\begin{figure}[t]
  \centering
  \begin{tabular}[h]{@{}lrrrrr@{}}
    & C code & Auto annot.\ & Manual annot.\ & Manual/loc 
    & VCC time
    \\
    Program & (loc) & (loc) & (loc) & (\%) 
    & (s)
    \\\hline
    Finite differences~\cite{foster} & 256 & 69 & 12 & 4.7 
    & 6.1
    \\
    Parallel dot product~\cite{pacheco} & 357 & 81 & 14 & 3.9 
    & 3.7
    \\
    Parallel Jacobi~\cite{pacheco} &  429 & 34 & 18 & 4.2 
    & 11.4
    \\
    N-body simulation~\cite{using-mpi} & 362 & 80 & 16 & 4.4 
    & 7.0
  \end{tabular}
  \caption{Lines of code, annotations and running times for four
    textbook programs}
\label{fig:statistics}
\end{figure}


We have manually annotated C+MPI code taken from standard textbooks.
Even though all annotations were introduced by hand, we distinguish
those that can be automatically generated in principle (e.g., the loop
and choice annotations) from those that necessarily require the
programmer's intervention (e.g, initialising the \ccode{type_func}
function with the size of the problem, line 9 in
Figure~\ref{fig:running-example-annotated}; loop invariants, lines
18--19; and non-MPI function annotations).
In the table in Figure~\ref{fig:statistics} we summarise the number of
lines of code (loc) in the original program, the number of annotations
that can be potentially automated, the number of programmer
annotations,
the number of manual annotations per 100 lines of original C code, and
the average time VCC took to complete the verification on a Windows
machine with two Intel 2.66 GHz cores and 2~GB of RAM
\footnote{VCC seems to make use of only one core in a multi-core platform. 
We have also performed our tests on a virtualised platform and observed no difference in performance between virtual machines configured with 1 and 2 cores.}.

Remark the small number of manual annotations required to successfully
verify the code, in all cases below 5\% when compared to the total
number of lines of code. 
As described in the previous section, VCC analyses each C function
separately. Hence, a major source of annotations comes from
functions. Since currently we annotate functions by hand, we have
analysed the function related annotations on the manual
annotations. How many of such annotations are needed heavily depends
on the nature of C code involved; however a fair amount of functions
are to be expected on a well designed code. In the examples we tested,
we added between 2.4 (finite differences) and 6.0 (parallel Jacobi)
lines of annotations per function and in average.

In addition to the numbers above we must add approximately 800 
lines of annotated C header files to describe the datatypes, axioms, and the
contracts for the various MPI primitives.
The annotated header files and examples are
available from {\small\url{http://www.di.fc.ul.pt/~edrdo/ampi-0.1.zip}}.



\section{Conclusion and future work}
\label{sec:conclusion}

We have identified a framework for checking C+MPI code against a
protocol description language. Terms of this language, called
\emph{communication types}, describe the overall communication
structure of programs. Such a type is then translated into a datatype
term of VCC, a verification tool for the C programming language, in
the form of an \texttt{\small include} header file. This header,
together with a VCC-annotated MPI-library header file, are fed into
VCC that determines whether the code follows the protocol.
We used our framework to check four non-trivial examples (260--430
lines of code), with promising results. 

Much remains to be done.
Currently, communication types are manually translated into VCC 
terms. We are working on an Eclipse plugin that checks the
good formation of types (are variables declared? is the source and
destination of messages valid ranks?) and generates the corresponding
VCC term.

Currently, all annotations to C code are performed by hand. We are
working on a tool that, given C source code, automatically inserts the
required annotations on collective operations (loop and
choice). Another major source of handcrafted annotations are C
functions: for each function a pre- and a post-condition on the
current communication type (in the form of \lstinline|requires| and
\lstinline|ensures| predicates) needs to be manually introduced. We
believe that, under some conditions (MPI programs are usually not
recursive), these contracts can be automatically derived, given the
communication type.

The communication language is currently quite limited. Even though it
features a dependent function constructor, its usage in practice is
quite restrictive (it can only occur at the top level). The next version of our
language will allow using the construct anywhere in the type, will
feature a for-each construct allowing to describe protocols with a
variable number of processes (cf.~\cite{PMSTypes}), as well
dependencies between communication primitives so that types may refer
to values exchanged in previous communications. Finally, we plan to
address further MPI operations, including non-blocking primitives
(\lstinline|MPI_Isend|, \lstinline|MPI_Irecv|, and
\lstinline|MPI_Wait|).

Currently, there is a too close connection between a \commcode{loop}
type in a communication type and a \lstinline|for| or a
\lstinline|while| loop in C code. We would like to relax the
connection by using  iso- or equi-recursive techniques.
There is also the (deep) issue of checking whether all processes
effectively follow the same branch in a collective operation.
Finally, the theory of communication types need to be further developed.


\paragraph{Acknowledgements}

Work funded by EPRSC (EP/K011715/1, EP/K034413/1 and EP/G015635/1) and
FCT (PTDC/ EIA-CCO/122547/2010) projects, and by LaSIGE (PEst-OE/EEI/
UI0408/2011).  We thank Ernie Cohen for technical help on VCC and
Nobuko Yoshida for comments and discussions.


\bibliographystyle{eptcs}
\bibliography{paper}

\end{document}